\documentclass[twocolumn,showkeys,aps,prb,showpacs]{revtex4-1}
\UseRawInputEncoding
\usepackage{graphicx}
\usepackage[CJKbookmarks,dvipdfm,colorlinks,linkcolor=red,citecolor=blue]{hyperref}

\begin{document}

\title{Electric-field induced magnetic-anisotropy transformation to achieve spontaneous valley polarization}

\author{San-Dong Guo$^{1}$,  Xiao-Shu Guo$^{1}$,  Guang-Zhao Wang$^{2}$, Kai Cheng$^{1}$, and Yee-Sin Ang$^{3}$}
\affiliation{$^1$School of Electronic Engineering, Xi'an University of Posts and Telecommunications, Xi'an 710121, China}
\affiliation{$^2$Key Laboratory of Extraordinary Bond Engineering and Advanced Materials Technology of Chongqing, School of Electronic Information Engineering, Yangtze Normal University, Chongqing 408100, China}
\affiliation{$^3$Science, Mathematics and Technology (SMT), Singapore University of Technology and Design (SUTD), 8 Somapah Road, Singapore 487372, Singapore}
\begin{abstract}
Valleytronics has been widely investigated  for providing new degrees
of freedom to future information
coding and processing. Here, it is proposed that valley
polarization can be achieved by electric field induced magnetic anisotropy (MA) transformation. Through the first-principle calculations,  our idea  is illustrated by a concrete example of $\mathrm{VSi_2P_4}$ monolayer. The increasing electric field  can induce a transition of MA from in-plane to out-of-plane by changing  magnetic anisotropy energy (MAE) from negative to positive value, which is mainly due to increasing magnetocrystalline anisotropy (MCA) energy. The out-of-plane magnetization is  in favour of spontaneous valley polarization in $\mathrm{VSi_2P_4}$. Within considered  electric field range, $\mathrm{VSi_2P_4}$ is always ferromagnetic (FM) ground state.  In a certain range of electric field, the coexistence of semiconductor and out-of-plane magnetization makes $\mathrm{VSi_2P_4}$  become a true ferrovalley (FV) material.
 The anomalous valley Hall effect (AVHE) can be observed under in-plane and out-of-plane electrical field in $\mathrm{VSi_2P_4}$. Our works  pave the way to design the ferrovalley material  by electric field.

\end{abstract}
\keywords{Electric field, Magnetic anisotropy, Ferrovalley  ~~~~~~~~~~~~~Email:sandongyuwang@163.com}

\maketitle

\section{Introduction}
The manipulation of the carriers' properties is key for semiconductor technology. In addition to charge and
spin of carriers, the valley, as a new degree of freedom,  is characterized by a local energy extreme in
the conduction band or valence band.  To encode, store and process information,  two or more degenerate but inequivalent valley states   at
the inequivalent $k$ points  should exist in a material\cite{b1,b2}.
  Since  the  two-dimensional (2D) materials, especially 2D 2H-phase transition metal dichalcogenides (TMDCs), are booming, the valley-related filed is truly flourishing\cite{q8-1,q8-2,q8-3,q9-1,q9-2,q9-3,q9-4}.
Nevertheless,   realizing  the valley polarization is  indispensable to achieve valley application.
To induce valley polarization, many methods have been proposed, such as  optical pumping, magnetic field, magnetic
substrates and  magnetic doping\cite{q8-1,q8-2,q8-3,q9-1,q9-2,q9-3,q9-4}. However,  these methods have some disadvantages that
the  intrinsic  energy band structures and crystal structures can be destroyed.
Recently, FV  materials have been proposed, which possess intrinsic spontaneous valley polarization introduced by their intrinsic ferromagnetism\cite{q10}. The  FV  materials can overcome these shortcomings of the extrinsic
valley polarization materials. Many 2D materials have been predicted to have spontaneous valley polarization\cite{q10,q11,q12,q13,q13-1,q14,q14-0,q14-1,q14-2,q14-3,q15,q16,q17,q18,q19}.
\begin{figure}
  \includegraphics[width=8cm]{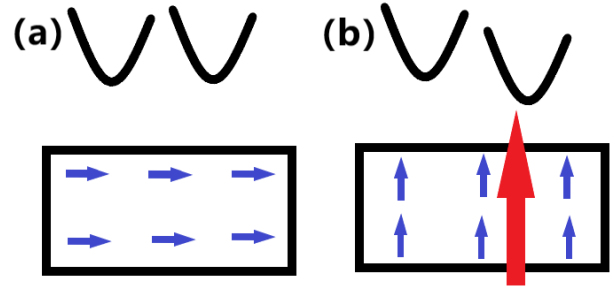}
  \caption{(Color online)(a): a 2D material with in-plane magnetization (blue arrows)  shows no valley polarization. (b): when an electric field (red arrow) is applied along out-of-plane direction, the  magnetization changes from in-plane to out-of-plane, and a valley polarization is induced. }\label{st}
\end{figure}

However, many of predicted 2D FV materials intrinsically have no spontaneous valley polarization, because their  magnetization are preferentially in the layer plane\cite{q14-1,q14-2,q14-3}. To achieve valley polarization in these materials, the external magnetic field is needed to adjust magnetization from in-plane to out-of-plane. Even though predicted 2D FV material possesses out-of-plane magnetization, the MAE is calculated with only considering MCA, not including magnetic shape anisotropy (MSA). For example, the  easy magnetization axis of $\mathrm{VSi_2P_4}$  is in-plane  orientation with including MSA\cite{gsd},  which  is different from previous out-of-plane one  with only considering MCA energy\cite{q14}.  Thus, MSA is very important to determine MA of 2D materials with weak spin-orbital coupling (SOC), like  the representative  2D magnet $\mathrm{CrCl_3}$ with  an easy in-plane magnetization in experiment\cite{a1-3,a1-4}.
 Only considering MCA energy shows an
easy out-of-plane magnetization for $\mathrm{CrCl_3}$, but  the theoretical results with including MSA energy agree well with  experimental in-plane magnetization\cite{a1-7}.

In this work, we   propose a  way
to achieve MA transformation from in-plane to out-of-plane by electric field, and then  realize  valley polarization (see \autoref{st}).
In \autoref{st} (a), if a 2D material is with in-plane magnetization,  no valley polarization can be observed. In \autoref{st} (b), when an electric field  is imposed
along out-of-plane direction, the  magnetization changes from in-plane to out-of-plane, and a valley polarization can be  induced.
Recently, 2D $\mathrm{MA_2Z_4}$
family  with a septuple-atomic-layer structure  has been established, and $\mathrm{MoSi_2N_4}$ and $\mathrm{WSi_2N_4}$ of them
  have been
synthesized  by the chemical vapor deposition
method\cite{a11,a12}.  Here, by the first-principle calculations, we use  a concrete example of  $\mathrm{VSi_2P_4}$ monolayer to illustrate our idea.
Calculated results show that increasing electric filed indeed can make MA change from in-plane to out-of-plane direction in $\mathrm{VSi_2P_4}$.
The semiconductor properties  and out-of-plane magnetization can coexist in a certain  electric field region, implying that $\mathrm{VSi_2P_4}$ becomes a true FV material. Our findings can be extended to other 2D materials, and tune MA by electric field.

\begin{figure}
  \includegraphics[width=8cm]{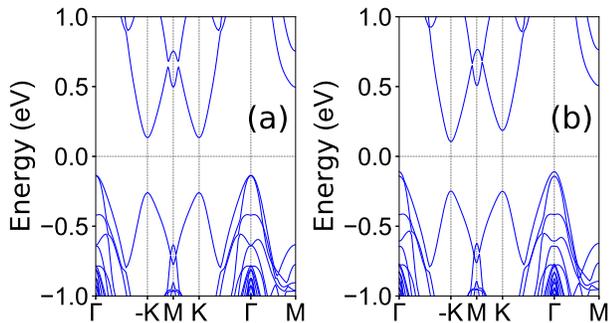}
  \caption{(Color online) For $\mathrm{VSi_2P_4}$ monolayer, the energy band structures with in-plane (a) and out-of-plane (b) without applying electric field. }\label{band}
\end{figure}
\begin{figure}
  \includegraphics[width=8cm]{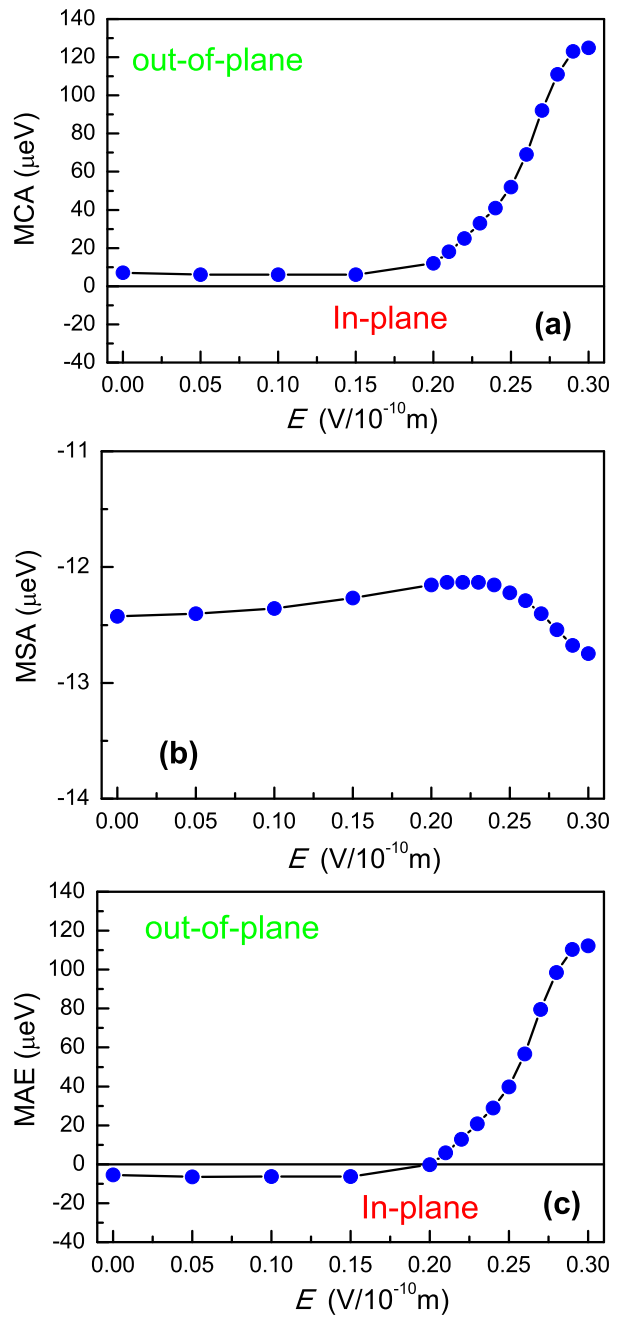}
  \caption{(Color online) For $\mathrm{VSi_2P_4}$ monolayer, MCA energy (a), MSA energy (b) and MAE (c) as a function of $E$. }\label{mae}
\end{figure}

\section{Computational detail}
Based on density-functional
theory  (DFT)\cite{1}, we perform the spin-polarized  first-principles calculations  by employing the projected
augmented wave method,  as implemented in Vienna ab initio simulation package (VASP)\cite{pv1,pv2,pv3}.
The generalized gradient approximation of Perdew-Burke-Ernzerhof (PBE-GGA)\cite{pbe} is adopted as exchange-correlation functional.
To consider on-site Coulomb correlation of V  atoms,  the GGA+$U$   method   in terms of  the on-site Coulomb
interaction of $U$$=$ 3.0 eV\cite{q14} is used within  the rotationally invariant approach proposed by Dudarev et al\cite{u}, where only the effective
$U$ ($U_{eff}$) based on the difference between the on-site Coulomb interaction
parameter  and exchange parameters  is meaningful.
The energy cut-off of 500 eV,  total energy  convergence criterion of  $10^{-8}$ eV and  force
convergence criteria of less than 0.0001 $\mathrm{eV.{\AA}^{-1}}$ on each atom are adopted to attain accurate results.
To avoid the interactions
between the neighboring slabs, a vacuum space of more than 32 $\mathrm{{\AA}}$ is used.
 The $\Gamma$-centered 16 $\times$16$\times$1 k-point meshs  in  the Brillouin zone (BZ) are used for structure optimization and electronic structures calculations, and 9$\times$16$\times$1 Monkhorst-Pack k-point meshs for calculating FM/antiferromagnetic (AFM)  energy  with rectangle supercell.
 The SOC effect is explicitly included to investigate MCA and  electronic  properties of  $\mathrm{VSi_2P_4}$ monolayer.
We calculate the Berry curvatures   directly from  wave functions  based on Fukui's
method\cite{bm} by using  VASPBERRY code\cite{bm1,bm2}.

\begin{figure}
  \includegraphics[width=8cm]{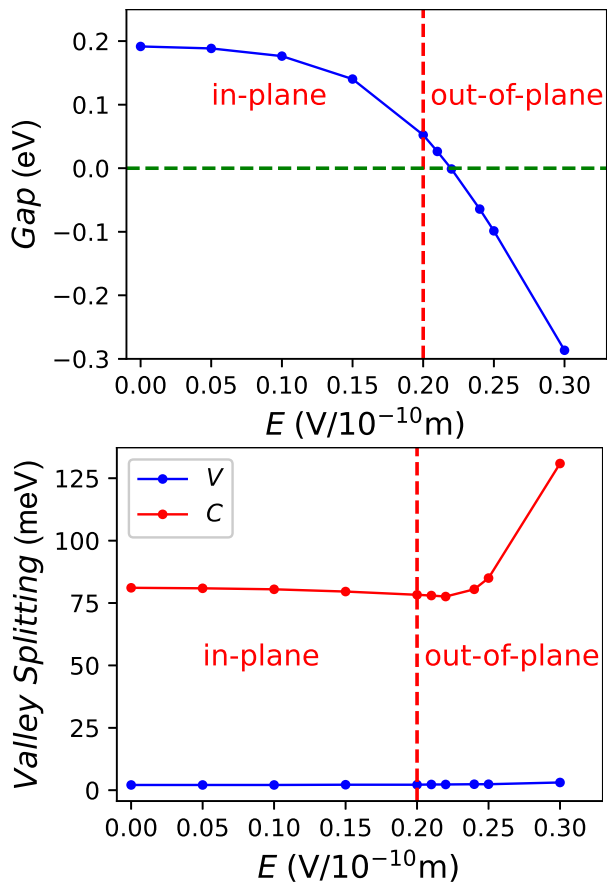}
  \caption{(Color online)For $\mathrm{VSi_2P_4}$ monolayer with fixed out-of-plane  MA,   the  global energy band gap (Top panel) and   the valley splitting  for both valence and condition bands (Bottom panel)  as a function of   $E$. The red vertical dashed line distinguishes actual in-pane and out-of-plane. }\label{gap}
\end{figure}

\begin{figure*}
  \includegraphics[width=16cm]{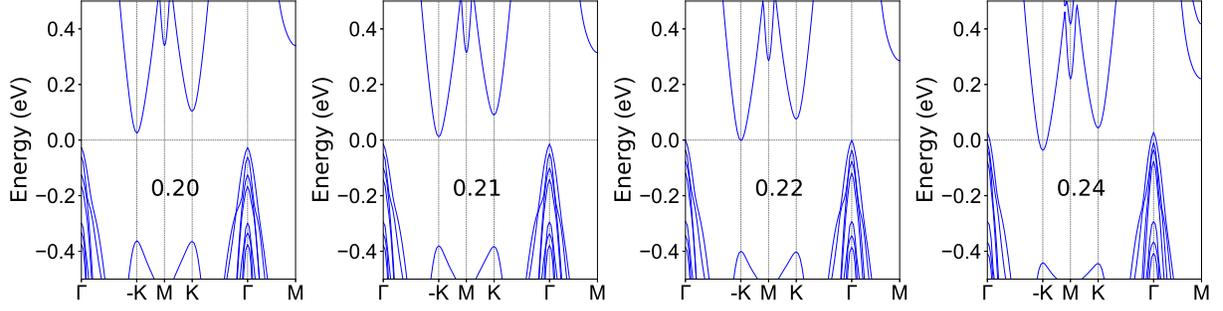}
  \caption{(Color online)For $\mathrm{VSi_2P_4}$ monolayer with fixed out-of-plane  MA, the energy band structures at representative $E$=0.20, 0.21, 0.22 and 0.24 $\mathrm{V/10^{-10}m}$. }\label{band-1}
\end{figure*}

\section{Structure and magnetic anisotropy}
The top and side views of  crystal structure of $\mathrm{VSi_2P_4}$
monolayer are shown in FIG.1 of electronic supplementary information (ESI), along with  the first BZ with high-symmetry points.
The V, Si and P atoms are packed in a honeycomb lattice with a space group of $P\bar{6}m2$ (No.187), which has  broken inversion symmetry, allowing spontaneous valley polarization. This crystal is composed of covalently bonded atomic layers in the order of
P-Si-P-V-P-Si-P along the $z$ axis. In other words,  the monolayer $\mathrm{VSi_2P_4}$ can be viewed as a $\mathrm{MoS_2}$-like
$\mathrm{VP_2}$ layer sandwiched in-between two slightly buckled
honeycomb SiP layers.
The optimized  lattice constants $a$ of $\mathrm{VSi_2P_4}$ (3.486 $\mathrm{{\AA}}$) is good
agreement with a previous report\cite{q14}.

 Because the magnetization is a pseudovector,  the out-of-plane FM  breaks all
possible vertical mirror symmetry, but  preserves the horizontal mirror symmetry, which allows the spontaneous valley polarization and a nonvanishing Chern number of  2D system.  The $\mathrm{VSi_2P_4}$ is predicted to be an out-of-plane ferromagnet (a FV material)  with  only considering MCA energy\cite{q14,gsd}. However, the  $\mathrm{VSi_2P_4}$  will become a 2D-$XY$ magnet (a non-FV material), when including MSA energy\cite{gsd}. The energy band structures with both in-plane and out-of-plane magnetization are plotted in \autoref{band}. It is clearly seen that $\mathrm{VSi_2P_4}$ is a FV material with out-of-plane  magnetization, and it will become a common magnetic semiconductor with in-plane case. Thus, just flipping the magnetic anisotropy through the external field, $\mathrm{VSi_2P_4}$ will become a FV material. Next, we will flip the magnetic anisotropy from in-plane to out-of-plane through  electric field.
\begin{figure}
  \includegraphics[width=8cm]{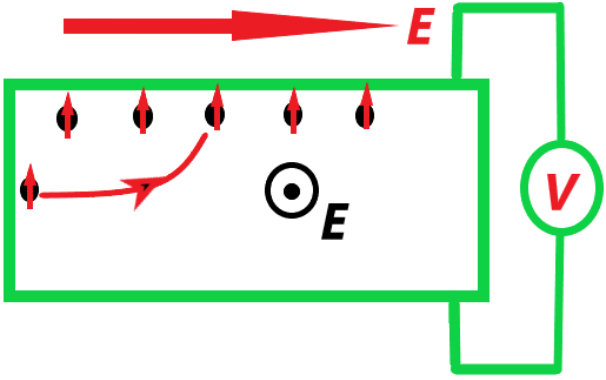}
  \caption{(Color online)For $\mathrm{VSi_2P_4}$ monolayer, schematic diagram of the AVHE along with the Hall voltage under an in-plane longitudinal electric field $E$.  The out-of-plane electric field $E$ is used to induce  magnetization change from in-plane to out-of-plane.  Upward small red arrows   represent spin-up carriers.  }\label{avh}
\end{figure}

Firstly, the magnetic ground state under the electric field is confirmed, and the energy differences (per formula unit) between AFM and FM ordering as a function of electric field $E$ are  plotted in FIG.2 of ESI.
It is found that the FM state is always  ground state within  considered $E$ range, and increasing $E$ can reduce the FM interaction.

The  orientation of magnetization can affect  magnetic, electronic  and topological properties of  2D materials\cite{q14-0,a4,a6,a7}.
The  orientation of magnetization can be  determined by MAE, which mainly includes
 two parts: (1) MCA energy $E_{MCA}$ caused by
the SOC; (2) MSA energy ($E_{MSA}$) due to
the anisotropic dipole-dipole (D-D) interaction\cite{a1-7,a7-1}:
 \begin{equation}\label{d-d}
E_{D-D}=\frac{1}{2}\frac{\mu_0}{4\pi}\sum_{i\neq j}\frac{1}{r_{ij}^3}[\vec{M_i}\cdot\vec{M_j}-\frac{3}{r_{ij}^2}(\vec{M_i}\cdot\vec{r_{ij}})(\vec{M_j}\cdot\vec{r_{ij}})]
 \end{equation}
in which  the $\vec{M_i}$ and $\vec{r_{ij}}$ represent the local magnetic moments and  vectors that connect the sites $i$ and $j$.
 For  monolayer with a collinear FM ordering, the $E_{MSA}$  ($E_{D-D}^{||}-E_{D-D}^{\perp}$) can be  expressed as:
\begin{equation}\label{d-d-3}
E_{MSA}=\frac{3}{2}\frac{\mu_0M^2}{4\pi}\sum_{i\neq j}\frac{1}{r_{ij}^3}\cos^2\theta_{ij}
 \end{equation}
where $||$ and $\perp$  mean that spins lie in
the plane and out-of-plane, and $\theta_{ij}$ is the angle between the $\vec{M}$ and $\vec{r_{ij}}$. The $E_{MSA}$ depends on the crystal structure and local magnetic moment. To minimize
magnetostatic energy,  the MSA tends to make spins directed parallel to the monolayer.
Usually, the MSA energy can be ignored, but  MSA becomes significant to MAE for
materials with weak SOC.

Without inclusion of SOC, there is no link between the crystalline structure
and the direction of the magnetic moments. Thus, the MCA energy is calculated  by $E_{MCA}=E^{||}_{SOC}-E^{\perp}_{SOC}$ from
GGA+$U$+SOC calculations. The calculation includes  two steps: (i) a collinear self-consistent field calculation without SOC to obtain the convergent charge density; (ii) a noncollinear non-self-consistent calculation of two different magnetization directions (The magnetization directions are in the $xy$ plane ($E^{||}_{SOC}$)
and along the $z$ axis ($E^{\perp}_{SOC}$)) within SOC. The negative/positive MAE corresponds to an easy axis
along the in-plane/out-of-plane direction.
We show $E_{MCA}$ as a function of $E$ in \autoref{mae} (a). Calculated results show  that  the $E_{MCA}$ is positive within considered $E$ range.
When $E$$<$0.2 $\mathrm{V/10^{-10}m}$, the $E_{MCA}$ basically remains unchanged with about 6 $\mathrm{\mu eV}$. For  $E$$>$0.2 $\mathrm{V/10^{-10}m}$, the $E_{MCA}$ has a rapid increase, and reaches to 125 $\mathrm{\mu eV}$ at $E$$=$0.3 $\mathrm{V/10^{-10}m}$. This implies that electric filed can effectively tune $E_{MCA}$.
The $E_{MSA}$ is calculated by \autoref{d-d-3}.  The local magnetic moment of V atom ($M_V$) as a function of $E$ is plotted  in FIG.3 of ESI. It is found that $M_V$ has small change within considered $E$ range (1.08 $\mu_B$-1.11 $\mu_B$). The $E_{MSA}$ versus $E$ is plotted in \autoref{mae} (b), and the $E_{MSA}$ changes from -12.13 $\mathrm{\mu eV}$ to -12.75 $\mathrm{\mu eV}$.  The MAE is calculated by  $E_{MAE}$=$E_{MCA}$+$E_{MSA}$, as shown in  \autoref{mae} (c).
When $U$ is less than 0.2 $\mathrm{V/10^{-10}m}$, the in-plane anisotropy can be observed due to negative MAE.  When $E$$>$0.2 $\mathrm{V/10^{-10}m}$, the positive MAE means a preferred out-of-plane magnetization. The MAE changes from negative value to positive value, implying a transition of MA from in-plane to out-of-plane induced by electric field.

\section{electronic structures}
Electric field can
result in semiconductor to metal transition  in bilayer $\mathrm{MoSi_2N_4}$ and $\mathrm{WSi_2N_4}$\cite{apl}. Thus, the electronic structures of $\mathrm{VSi_2P_4}$ under electric field  are investigated to confirm its semiconductor properties. With fixed out-of-plane magnetization, the evolutions of total energy band gap  and the valley splitting for both valence and condition bands  as a function of $E$ are plotted in \autoref{gap}.  At representative $E$, the energy band structures are plotted in \autoref{band-1}. It is found that the gap decreases with increasing $E$, and a semiconductor to metal transition is produced at $E$$=$0.22  $\mathrm{V/10^{-10}m}$.  Combined with the previous out-of-plane MA $E$ region ($E$$>$0.20 $\mathrm{V/10^{-10}m}$), the $\mathrm{VSi_2P_4}$  is intrinsically a FV material for $E$ between 0.20 $\mathrm{V/10^{-10}m}$ and 0.22 $\mathrm{V/10^{-10}m}$.

Within considered $E$ range,  the valley splitting of conduction band is observable, while the valley splitting of valence band can be ignored. This can be explained by distribution of V-$d$ orbitals.   The  V-$d$ orbitals split
into  $d_z^2$ orbital, $d_{xy}$+$d_{x^2-y^2}$  and
$d_{xz}$+$d_{yz}$ orbitals in a trigonal prismatic crystal field environment. According to FIG.4 of ESI, the $d_{x^2-y^2}$+$d_{xy}$ orbitals dominate K and -K valleys of  conduction  bands, while the $d_{z^2}$ orbitals mainly contribute to K and -K valleys of  valence  bands.
Since the FM ordering  breaks spin degeneracy between
the spin-up and spin-down bands, the SOC Hamiltonian only
involving  the interaction of the same spin states mainly produces valley polarization.
If $d_{x^2-y^2}$+$d_{xy}$  orbitals dominate  the K and -K valleys,  the valley splitting $|\Delta E|$  can be expressed as\cite{v2,v3}:
\begin{equation}\label{m4}
|\Delta E|=|E^{K}-E^{-K}|=4\alpha
\end{equation}
If the -K and K valleys are mainly from $d_{z^2}$ orbitals, the valley splitting $|\Delta E|$ is given:
\begin{equation}\label{m4}
|\Delta E|=|E^{K}-E^{-K}|=0
\end{equation}
where $\alpha$ and $E^{K}$/$E^{-K}$ are the SOC-related constant and  the resulting energy level at K/-K valley. For $d_{x^2-y^2}$+$d_{xy}$-dominated -K and K valley with general magnetization orientation,   $|\Delta E|$$=$$4\alpha cos\theta$\cite{v3}, where $\theta$=0/90$^{\circ}$ for out-of-plane/in-plane magnetization.  Thus,  the valley splitting of $\mathrm{VSi_2P_4}$ for both valence and conduction bands will  become zero, when $\mathrm{VSi_2P_4}$ is with in-pane case.

FIG.5 and FIG.6 of ESI  present the calculated energy band structures and  Berry curvatures of
$\mathrm{VSi_2P_4}$ under $E$$=$0.21  $\mathrm{V/10^{-10}m}$ with SOC for magnetic moment of V along the $+z$ and  $-z$ directions. FIG.5 (a) shows  the  valley polarization, and the energy of K valley
is higher than one of -K valley.  The valley polarization can  be
switched by reversing the magnetization direction from $+z$ to $-z$ direction (see FIG.5 (b) of ESI).
The  $\mathrm{VSi_2P_4}$ under $E$$=$0.21  $\mathrm{V/10^{-10}m}$ is  an indirect band
gap semiconductor with gap value of 26.4 meV.
For the two situations, the opposite
signs of Berry curvature around -K and K valleys can be observed. It is found that there are the unequal magnitudes of Berry curvatures at -K and K valleys.   By reversing
the magnetization, the magnitudes
of Berry curvature at -K and K valleys exchange to each
other,   but their signs remain unchanged.

Under an in-plane longitudinal  $E$, anomalous velocity $\upsilon$  of Bloch electrons at K and -K valleys  is related with Berry curvature $\Omega(k)$:$\upsilon\sim E\times\Omega(k)$\cite{qqq}.
An appropriate electron doping
makes the Fermi level  fall between the -K and K
valleys of conduction band.   When  in-plane and out-of-plane  electric fields are applied, the  Berry curvature forces
the electron carriers to accumulate on one side of the sample,  giving rise to an
AVHE in monolayer $\mathrm{VSi_2P_4}$ (see \autoref{avh}).  In fact, for example, when out-of-plane  electric field  $E$$=$0.24  $\mathrm{V/10^{-10}m}$,
only -K valley of conduction band is conductive (see \autoref{band-1}), and AVHE can be achieved with applied electric fields.

\section{Conclusion}
In summary, we have demonstrated that the electric field  can result in a MA transformation
in $\mathrm{VSi_2P_4}$.  The electric field can induce sudden change of positive MCA energy, and has small effects on negative MSA energy.
This leads to sign change of MAE, giving rise to in-plane to out-of-plane transformation of MA. It is found that electric field can induce semiconductor-metal phase transition in $\mathrm{VSi_2P_4}$. However, for a certain electric field region, semiconductor properties  and out-of-plane magnetization can coexist, confirming FV properties of $\mathrm{VSi_2P_4}$.  When  in-plane and out-of-plane  electric fields are applied, an
AVHE in monolayer $\mathrm{VSi_2P_4}$ can be achieved.
Our findings can inspire more works of electric filed-tuned MA.
~~~~\\
~~~~\\
\textbf{SUPPLEMENTARY MATERIAL}
\\
See the supplementary material for  crystal structures; energy difference between AFM and FM and local magnetic moment of V atom as a function of $E$; the related energy band structures and Berry curvatures.

~~~~\\
~~~~\\
\textbf{Conflicts of interest}
\\
There are no conflicts to declare.

\begin{acknowledgments}
This work is supported by Natural Science Basis Research Plan in Shaanxi Province of China  (2021JM-456). We are grateful to Shanxi Supercomputing Center of China, and the calculations were performed on TianHe-2.
\end{acknowledgments}


\begin{references}
\bibitem{b1}L. J. Sham, S. J. Allen, A. Kamgar, and D. C. Tsui, Phys. Rev.
Lett. \textbf{40}, 472 (1978).


\bibitem{b2} S. A. Wolf, D. D. Awschalom, R. A. Buhrman, J. M. Daughton,
S. von Moln¨¢r, M. L. Roukes, A. Y. Chtchelkanova, and D. M.
Treger, Science \textbf{294}, 1488 (2001).

\bibitem{q8-1} H. Zeng, J. Dai, W. Yao, D. Xiao, and X. Cui, Nat. Nanotechnol.
\textbf{7}, 490 (2012).

\bibitem{q8-2} K. F. Mak, K. He, J. Shan, and T. F. Heinz, Nat. Nanotechnol.
\textbf{7}, 494 (2012).


\bibitem{q8-3}A. Srivastava, M. Sidler, A. V. Allain, D. S. Lembke, A. Kis,
and A. Imamo¡äglu, Nat. Phys. \textbf{11}, 141 (2015).

\bibitem{q9-1}8 H. Zeng, J. Dai, W. Yao, D. Xiao and X. Cui, Nat.
Nanotechnol. \textbf{7}, 490 (2012).


\bibitem{q9-2}D. MacNeill, C. Heikes, K. F. Mak, Z. Anderson,
A. Korm$\acute{a}$nyos, V. Z$\acute{o}$lyomi, J. Park and D. C. Ralph, Phys.
Rev. Lett. \textbf{114}, 037401 (2015).



\bibitem{q9-3} C. Zhao, T. Norden, P. Zhang, P. Zhao, Y. Cheng, F. Sun,
J. P. Parry, P. Taheri, J. Wang, Y. Yang, T. Scrace, K. Kang,
S. Yang, G. Miao, R. Sabirianov, G. Kioseoglou, W. Huang,
A. Petrou and H. Zeng, Nat. Nanotechnol.  \textbf{12}, 757 (2017).


\bibitem{q9-4} M. Zeng, Y. Xiao, J. Liu, K. Yang and L. Fu, Chem. Rev. \textbf{118}, 6236  (2018).


\bibitem{q10}W. Y. Tong, S. J. Gong, X. Wan, and C. G. Duan,
Nat. Commun. \textbf{7}, 13612 (2016).

\bibitem{q11}Y. B. Liu, T. Zhang, K. Y. Dou, W. H. Du, R. Peng, Y. Dai, B. B. Huang,
and Y. D. Ma, J. Phys. Chem. Lett. \textbf{12}, 8341 (2021).

\bibitem{q12}Z. Song, X. Sun, J. Zheng, F. Pan, Y. Hou, M.-H. Yung, J. Yang,
and J. Lu, Nanoscale \textbf{10}, 13986 (2018).


\bibitem{q13}J. Zhou, Y. P. Feng, and L. Shen, Phys. Rev. B \textbf{102}, 180407(R)
(2020).


\bibitem{q13-1}P. Zhao, Y. Ma, C. Lei, H. Wang, B. Huang, and Y. Dai,
Appl. Phys. Lett. \textbf{115}, 261605 (2019).



\bibitem{q14}X. Y. Feng, X. L. Xu, Z. L. He, R. Peng, Y. Dai, B. B. Huang and Y. D. Ma,  Phys. Rev. B \textbf{104}, 075421 (2021).
\bibitem{q14-0}S. Li, Q. Q. Wang, C. M. Zhang, P. Guo and S. A. Yang,  Phys. Rev. B  \textbf{104}, 085149 (2021).

\bibitem{q14-1}H. X. Cheng, J. Zhou, W. Ji, Y. N. Zhang and Y. P. Feng, Phys. Rev. B \textbf{103}, 125121 (2021).


\bibitem{q14-2}K. Sheng, Q. Chen, H. K. Yuan and Z. Y. Wang, Phys. Rev. B \textbf{105}, 075304 (2022)).


\bibitem{q14-3}P. Jiang, L. L. Kang,  Y. L. Li,  X. H. Zheng,  Z. Zeng  and S. Sanvito,  Phys. Rev. B \textbf{104}, 035430 (2021).





\bibitem{q15}Y. Zang, Y. Ma, R. Peng, H. Wang, B. Huang, and Y. Dai,
Nano Res. \textbf{14}, 834 (2021).



\bibitem{q16}R. Peng, Y. Ma, X. Xu, Z. He, B. Huang, and Y. Dai, Phys. Rev.
B \textbf{102}, 035412 (2020).


\bibitem{q17}W. Du, Y. Ma, R. Peng, H. Wang, B. Huang, and Y. Dai,
J. Mater. Chem. C \textbf{8}, 13220 (2020).

\bibitem{q18}R. Li, J. W. Jiang, W. B. Mi  and H. L. Bai, Nanoscale  \textbf{13}, 14807 (2021).


\bibitem{q19}S. D. Guo, J. X. Zhu, W. Q. Mu and B. G. Liu, Phys. Rev. B \textbf{104}, 224428 (2021).


\bibitem{gsd}S. D. Guo,  Y. L. Tao,  K. Cheng, B. Wang and Y.-S. Ang, arXiv:2207.13420 (2022).

\bibitem{a1-3}Z. Wang, M. Gibertini, D. Dumcenco, T. Taniguchi, K.
Watanabe, E. Giannini, and A. F. Morpurgo, Nat. Nanotechnol.
\textbf{14}, 1116 (2019).

\bibitem{a1-4}D. R. Klein, D. MacNeill, Q. Song, D. T. Larson, S. Fang, M.
Xu, R. A. Ribeiro, P. C. Canfield, E. Kaxiras, R. Comin, and
J. H. Pablo, Nat. Phys. \textbf{15}, 1255 (2019).


\bibitem{a1-7}X. B. Lu, R. X. Fei, L. H. Zhu and  L. Yang, Nat. Commun. \textbf{11}, 4724 (2020).

\bibitem{a11}Y. L. Hong, Z. B.  Liu, L. Wang  T. Y. Zhou,  W. Ma, C. Xu, S. Feng,
L. Chen, M. L. Chen, D. M. Sun, X. Q. Chen, H. M. Cheng and W. C. Ren, Science  \textbf{369}, 670 (2020).

\bibitem{a12}L. Wang, Y. Shi, M. Liu, A. Zhang, Y.-L. Hong, R. Li,
Q. Gao, M. Chen, W. Ren, H.-M. Cheng, Y. Li, and X.-
Q. Chen, Nature Communications \textbf{12}, 2361 (2021).






\bibitem{1}P. Hohenberg and W. Kohn, Phys. Rev. \textbf{136},
B864 (1964); W. Kohn and L. J. Sham, Phys. Rev. \textbf{140},
A1133 (1965).



\bibitem{pv1} G. Kresse, J. Non-Cryst. Solids \textbf{193}, 222 (1995).

\bibitem{pv2} G. Kresse and J. Furthm$\ddot{u}$ller, Comput. Mater. Sci. 6, \textbf{15} (1996).

\bibitem{pv3} G. Kresse and D. Joubert, Phys. Rev. B \textbf{59}, 1758 (1999).
\bibitem{pbe}J. P. Perdew, K. Burke and M. Ernzerhof, Phys. Rev. Lett. \textbf{77}, 3865 (1996).


\bibitem{u}S. L. Dudarev, G. A. Botton, S. Y. Savrasov, C. J. Humphreys and A. P. Sutton, Phys. Rev. B \textbf{57}, 1505 (1998).



\bibitem{bm}T. Fukui, Y. Hatsugai and H. Suzuki,  J. Phys. Soc. Japan. \textbf{74},
1674 (2005).


\bibitem{bm1}H. J. Kim,  https://github.com/Infant83/VASPBERRY, (2018).
\bibitem{bm2}H. J. Kim, C. Li, J. Feng, J.-H. Cho, and Z. Zhang, Phys. Rev. B  \textbf{93}, 041404(R) (2016).


\bibitem{a4}X. Liu, H. C. Hsu, and C. X. Liu, Phys. Rev. Lett. \textbf{111}, 086802 (2013).



\bibitem{a6}S. D. Guo, J. X. Zhu, M. Y. Yin and B. G. Liu, Phys. Rev. B  \textbf{105}, 104416 (2022).

\bibitem{a7}S. D. Guo, W. Q. Mu and B. G. Liu, 2D Mater.  \textbf{9}, 035011 (2022).






\bibitem{a7-1}K. Yang, G. Y. Wang, L. Liu ,  D. Lu  and H. Wu, Phys. Rev. B  \textbf{104}, 144416 (2021).


\bibitem{apl}Q. Y. Wu,  L. M. Cao,  Y. S. Ang and  L. K. Ang, Appl. Phys. Lett. \textbf{118}, 113102 (2021).


\bibitem{v2}P. Zhao, Y. Dai, H. Wang, B. B. Huang and  Y. D. Ma, ChemPhysMater, \textbf{1}, 56 (2022).

\bibitem{v3}R. Li, J. W. Jiang, W. B. Mi  and H. L. Bai, Nanoscale  \textbf{13}, 14807 (2021).


\bibitem{qqq}D. Xiao, M. C. Chang, and Q. Niu, Rev. Mod. Phys. \textbf{82}, 1959
(2010).

\end{references}
\end{document}